\documentstyle[epsf,aps,prb]{revtex}

\begin{document}
\draft

\twocolumn[\hsize\textwidth\columnwidth\hsize\csname
@twocolumnfalse\endcsname

\title{4$f$ spin density in the reentrant ferromagnet $\rm
SmMn_{2}Ge_{2}$}

\author{J.E. McCarthy$^1$, J.A. Duffy$^2$\cite{JAD}, C. Detlefs$^1$,
M.J. Cooper$^2$, P. Canfield$^3$ }

\address{ 
$\rm ^1$European Synchrotron Radiation Facility, BP 220,
F-38043 Grenoble, France\\
$\rm ^2$Department of Physics, University of   
Warwick, Coventry CV4 7AL, UK \\
$\rm ^3$Ames Laboratory, 59 Physics,
Iowa State University, Ames, Iowa 50011, USA\\ }

\date{\today}

\maketitle

\widetext

\begin{abstract}
The spin contribution to the magnetic moment in $\rm SmMn_{2}Ge_{2}$
has been measured by magnetic Compton scattering in both the low and
high temperature ferromagnetic phases. At low temperature, the Sm site
is shown to possess a large 4$f$ spin moment of $3.4 \pm 0.1 \mu_{B}$,
aligned antiparallel to the total magnetic moment. At high
temperature, the data show conclusively that ordered magnetic moments
are present on the samarium site.
\end{abstract} 

\vskip2pc]

\narrowtext 

The ternary compounds of structure $RT_2X_2$, where $R$ is a rare
earth, $T$ is a transition metal and $X$ is either Si or Ge, are of
considerable interest since they exhibit a wide variety of phenomena,
ranging from heavy fermion behavior and superconductivity to strong
ferro- and antiferromagnetism\cite{szytula:94,szytula:89}. The $R \rm
Mn_2Ge_2$ sub-series is of particular interest as the transition metal
carries a magnetic moment. These compounds crystallize in the
ThCr$_2$Si$_2$-type body-centered tetragonal structure consisting of
layers stacked along the $c$-axis in the sequence $R \rm - Ge - Mn_2 -
Ge - \it R$ (Ref.~\onlinecite{szytula:94,szytula:89}). The magnetic
ordering depends critically on the planar Mn--Mn distance, $d$. For
values greater than $d_c = 2.87 \rm \AA$ ferromagnetism is observed,
whereas below this antiferromagnetic ordering is favored. In $\rm
SmMn_{2}Ge_{2}$, $d$ is approximately equal to this critical value
\cite{sampa:96}, and numerous magnetization measurements
\cite{fujii:85,gyorgy:87} have shown a complex temperature dependence
of the magnetic ordering in the compound. Furthermore, this material
exhibits giant magnetoresistance of magnitude $\approx 8\%$ associated
with the antiferromagnetic phase\cite{dover:93}.  As a
naturally-layered material, the properties of $\rm SmMn_{2}Ge_{2}$
provide an interesting complement to studies of artificial multilayer
materials.

The purpose of the present study was to determine the moments on the
Sm site at different temperatures. The spin moments on the Sm and Mn
sites were determined by fitting atomic models to the magnetic Compton
profiles (MCP). 

One of the key objectives of earlier work has been to determine the
different magnetic structures of this material at various temperatures
\cite{sampa:96,fujii:85,gyorgy:87,lord:96,tomka:97,tomka:98}.  $\rm
SmMn_{2}Ge_{2}$ has three magnetically ordered phases. At 345K (above
which it is paramagnetic), the material becomes ferromagnetic. Below
155K, antiferromagnetic ordering occurs, and remains for temperatures
down to 105K (Ref.~\onlinecite{tomka:98}), then the compound becomes
ferromagnetic once more. The ordering is strongly anisotropic: in the
high temperature ferromagnetic phase, the easy axis lies along the
$[001]$ direction but along the $[110]$ direction in the low
temperature ferromagnetic phase. A further phase transition has been
suggested at $\rm T_c \approx 30K$ and was thought to arise from an
ordering of the Sm magnetic moment \cite{sampa:96}.

Very recently, Tomka {\it et al.} (Ref.~\onlinecite{tomka:98}) used
powder neutron diffraction to study the origin of the magnetic moments
in this material. They revealed a complex non-collinear buckled
structure with net ferromagnetic components of $\approx 2\mu_{B}$ per
Mn and up to $0.65\mu_{B}$ per Sm site. It was shown that the Mn
moments are not aligned with any high symmetry direction of the
crystal, and that the change in orientation of the easy magnetization axis
results from a change of the coupling, which leads to cancellation of
the net ferromagnetic moment components within the basal plane or
along $[001]$ in the high and low temperature phases,
respectively. These measurements were performed on isotopically
enriched samples in order to minimize the prohibitive neutron
absorption in natural samarium. Tomka {\it et al.} derived these
results from a Rietveld refinement which relies on tabulated values
for the neutron magnetic form factors. They assumed the Sm moment to
be induced by the Mn net ferromagnetic moment and therefore they
refined only its collinear projection. While Tomka {\it et al.} were
mostly concerned with the Mn antiferromagnetism, our current work
focuses on the distribution of the net ferromagnetic moments, mainly
on the Sm sublattice. For more details on the antiferromagnetic
structures, we refer the reader to Ref.~\onlinecite{tomka:98}.

Samarium has electronic configuration $4f^{5}$, and its Hund's rules
ground state is predicted to have a small total moment of
$0.84\mu_{B}$ arising from large antiparallel spin ($S = 5/2$) and
orbital ($L = 5$) angular momenta. However, another $J$-multiplet lies
just above this free-ion ground state, and crystalline electric field
(CEF) effects frequently lead to different ground states: hence Hund's
rules are expected to be unreliable in this system. Indeed in pure Sm
metal the paramagnetic moment is observed to be $1.5\mu_{B}$, instead
of the $0.84\mu_{B}$ expected. Consequently, knowledge of the spin and
orbital moments is particularly important in order to understand the
magnetization of this material. Magnetic form factors are sensitive to
the assumed $S$ and $L$ values of the magnetic ground state, leading
to uncertainties in the total Sm moment derived from neutron powder
data. X-ray magnetic circular dichroism (XMCD) is not an ideal
technique in this case because the validity of the sum rules is not
well established for the investigation of 4$f$ materials. This
magnetic Compton scattering (MCS) measurement, however, provides this
essential information directly and unambiguously.

MCS is a uniquely sensitive probe of the spin component of the
magnetization. The Compton effect is observed when high energy photons
are scattered off electrons. For bound electrons which have some
distribution of momenta, the scattered photon energy is Doppler
broadened into an energy distribution. The Compton profile,
$J(p{_z})$, is defined as the 1-dimensional projection of the electron
momentum distribution, $n({\bf p})$, onto the scattering vector, taken
to be parallel to the $z$ direction.

\begin{equation} 
J(p {_z} ) = \int \int n({\bf p}) {\rm d}p_{x} {\rm d} p_{y}.
\label{ccp} 
\end{equation} 

Within the impulse approximation \cite{platzman:70}, the Compton
profile is directly proportional to the scattering
cross-section\cite{holm:88}. The integral of $J(p{_z})$ is simply the
total number of electrons per unit cell. If the photons impinging on a
sample have a component of circular polarization, then a small spin
dependence appears in the scattering cross section\cite{bell:96}.
Reversing either the photon polarization or the magnetization of the
sample changes the sign of the spin-dependent signal, which enables
the spin part to be isolated. The resultant magnetic Compton profile
(MCP), is a projection of the momentum density of only those electrons
with unpaired spins,

\begin{equation} 
J_{\rm{mag}}(p {_z} ) = \int \int \left(n{^\uparrow}({\bf p}) -
n{^\downarrow}({\bf p}) \right) {\rm d}p_{x} {\rm d} p_{y}.
\label{mcp} 
\end{equation} 

Here, $n{^\uparrow}({\bf p})$ and $n{^\downarrow}({\bf p})$
are the momentum dependent majority and minority spin densities
respectively. The area under the MCP is equal to the number of
unpaired electrons, i.e. the total spin moment per formula unit in
Bohr magnetons.

Magnetic Compton scattering is now an established technique for
probing momentum space spin densities and band structures in magnetic
materials \cite{sakai:96,cooper:97}. Within the impulse approximation,
the method is solely sensitive to {\it spin} magnetic moments, $S$
(Ref. \onlinecite{sakai:96,cooper:92,lovesey:96}); the orbital moment,
$L$, is not measured \cite{carra:96}. This is especially useful in the
light rare earths and actinides, where $J= L-S$ may be small, even if
$L$ and $S$ are large. Unlike XMCD, MCS is equally sensitive to all
spin polarized electrons, regardless of their binding energy and the
symmetry of their wave functions. Since the MCP is a difference
between Compton profiles, the contributions from the spin-paired
electrons, and from unwanted systematic sources cancel out. In the
study of Sm and related materials, high energy x-rays have the
additional advantage that they do not suffer from the large absorption
factors associated with neutrons, which have severely hindered such
experiments \cite{tomka:98}.

The basal plane MCP for $\rm SmMn_{2}Ge_{2}$ was measured on the high
energy x-ray beamline, ID15, of the ESRF.  The experiment was
performed in reflection geometry \cite{duffy:98} with a scattering
angle of 168$^{\circ}$. The incident beam energy of 296keV was
selected by the $\{ 311 \}$ reflection of a Si monochromator. The
sample was grown by the slow cooling of a ternary melt rich in Mn and
Ge (Ref.~\onlinecite{canfield:92}). For the experiment a piece of
dimensions $5 \times 3 \times 1.5 {\rm mm}^3$ was cut from the
resultant crystal, and was oriented so that the resolved direction was
in the basal plane to within $\pm 2^{\circ}$.

The temperature of the sample was maintained at $15 \pm 2$K, $40 \pm
2$K and $230 \pm 2$K for the three measurements. The sample's
magnetization was reversed with a 0.96T rotating permanent magnet.
For the two low temperature measurements, this was sufficient to
saturate the magnetic moment, while at 230K the moment was
approximately $50\%$ saturated, since in the high temperature
ferromagnetic phase the easy axis is perpendicular to the basal
plane. A degree of circular polarization of $P_c \approx 45\%$ was
obtained by selecting a beam 20$\mu rad$ above the orbital plane of
the synchrotron. The energy spectrum of the scattered x-rays was
measured by a solid-state Ge detector. The momentum resolution
obtained was 0.44 atomic units (a.u., where 1 a.u. = $1.99 \times 10
^{-24}$ kg m s$^{-1}$). The total number of counts in each of the
charge profiles was $1.5 \times 10^{8}$, resulting in $3.7 \times
10^{6}$ in the MCP with a statistical precision of $\pm 3\%$ at the
magnetic Compton peak in a bin of width 0.09 a.u. The usual correction
procedures \cite{zuko:93} for the energy dependence of the detector
efficiency, absorption, the relativistic scattering cross-section and
magnetic multiple-scattering were applied, and after checking that the
profiles were symmetric about zero momentum, the MCPs were folded
about this point to increase the effective statistical precision of
the data. The amplitude of the MCP spectra, $J_{\rm{mag}}(p{_z})$, was
calibrated using data for Fe and Ni obtained under the same
experimental conditions to correct for the partial circular
polarization of the incident beam and other geometrical factors.

\begin{figure} 
\epsfxsize=220pt  
\epsffile{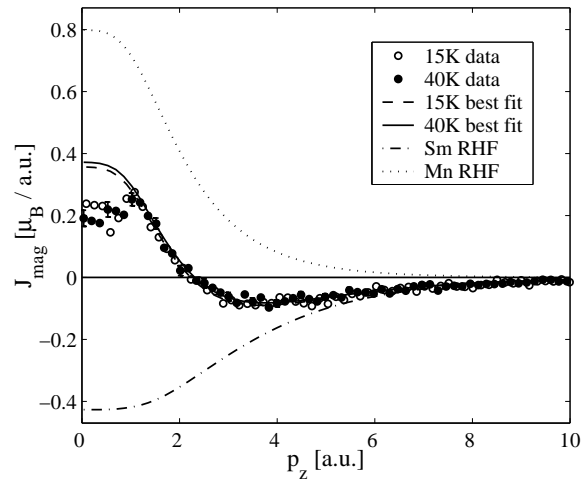} 
\hspace{1cm}
\caption{The experimental magnetic Compton profile of $\rm
SmMn_{2}Ge_{2}$ at $T = 15$K and $T = 40$K. The fits were performed
for $p_{z} > 1.5$a.u., using RHF predictions for the Mn 3$d$ and Sm
4$f$ moment, convoluted with a Gaussian of FWHM = 0.44 a.u. to
represent the experimental resolution.
\label{fig1} 
}
\end{figure}

The results from the measurements at 15K and 40K are shown in
Figure~\ref{fig1}, together with model profiles for Sm 4$f$ and Mn
3$d$ electrons based on relativistic Hartree Fock (RHF) free atom
wavefunctions \cite{biggs:75}. The model profiles have been scaled to
provide a best least-squares fit to the data for $p{_z} > 1.5$
a.u.. The fact that these free atom model profiles must provide an
accurate description of the Compton profiles at high momentum results
from energy considerations. The kinetic energy of the electron
distribution is given by the second moment of the Compton profile.
The virial theorem ensures that the kinetic energy and the total
energy are numerically equal but opposite in sign. Therefore the very
small total energy changes responsible for cohesion derive from
changes in the electron distribution at low momenta, associated with
electron density away from the atomic cores. The profiles for Mn 3$d$
and Sm 4$f$ electrons are significantly different, the latter being
$50\%$ broader, and therefore fitting at high momenta can be used to
separate the moments.  This difference can be thought of as simply
arising from the fact that the Sm 4$f$ electrons are more tightly
bound than the Mn 3$d$ electrons, and this difference manifests itself
in higher momentum components for the 4$f$ against the 3$d$ electrons:
a result which is also evident from simple consideration of the
uncertainty principle.

The analysis of magnetic Compton line shapes in terms of the
characteristically different orbital profiles has been demonstrated in
many similar cases e.g. HoFe$_2$ \cite{zuko:93}, CeFe$_2$
\cite{cooper:96} and UFe$_2$ \cite{lawson:97}.  Our results presented
in Figure~\ref{fig1} clearly show that there is a large negative 4$f$
spin moment, opposed to the positive Mn 3$d$ moment. Even though the
magnetic configuration of Sm is sensitive to the CEF environment, the
MCP of Sm and other 4$f$ materials indicate that deviations from the
atomic behavior are small, a result also supported by neutron
diffraction data.  Thus the area under the fitted 4$f$ curve gives a
reliable estimate of the Sm spin moment. Combination of bulk
magnetization data with the measured spin moments allow us to infer
the size of the orbital moment. These values, given in
Table~\ref{tab1} were determined as follows.

The total spin moment, calculated simply by integrating the MCP, is
effectively zero for the 15K and 40K data. The area under the Sm $4f$
profile at 40K was deduced to be $3.4 \pm 0.1 \mu_{B}$ per formula
unit. This means that the spin moment associated with the Mn $3d$
electrons together with the delocalised electrons, also amounts to the
same value, $3.4\mu_{B}$ per formula unit, but aligned
antiparallel. Note that the fitted Mn $3d$ profile (dotted line in
Figure~\ref{fig1}) is inappropriate at low momenta ($p{_z} < 1.5$a.u.)
because the $3d$ electrons are sensitive to the solid state
environment, and their contribution will differ from free atom
behavior, unlike the Sm $4f$ electrons. In addition small
contributions at low momentum from both delocalised Mn ($4sp$-like)
and Sm ($5d$- and $6sp$-like) electrons, are evident in
Figure~\ref{fig1} from the discrepancy between the data and the fitted
curve at low momentum. Similar effects are found in studies on other
$3d$ systems \cite{dixon:98}, and an electronic structure calculation
would be needed to examine this further. The main interpretation of
this MCP is that there is a Sm $4f$ spin moment of $3.4\mu_{B}$, which
is aligned antiparallel to the Mn $3d$, and total, magnetizations.
Since the total spin moment is zero, we can also deduce the size of
the orbital moment in this material. In order to account for the
macroscopic ferromagnetic moment of $4.1\mu_{B}$ measured by SQUID
magnetometry there must be an orbital moment of this size. This
orbital moment is aligned with the total magnetization, i.e. parallel
to the Mn spin direction. A large orbital contribution is expected
from the Sm $4f$ shell ($L=5$), while the orbital moment of
delocalised electrons is usually quenched. The transition metals
represent an intermediate case, where spin-orbit coupling and CEF
effects are of comparable strength, therefore a $3d$ orbital moment
will be reduced, but can not be ruled out completely, as has been
shown recently in NiO (Ref. \onlinecite{Fernandez}).  In the
Mn$^{2+}$ ion the $3d$ shell is exactly half filled ($S$=5/2, $L$=0),
so that the orbital moment vanishes. In $\rm SmMn_2Ge_2$, however, the
valence of Mn is not known, so that an orbital moment on the Mn site
cannot be neglected {\em a priori}. Nevertheless, the comparison of
MCP and bulk magnetization clearly shows that at low temperature there
is a significant contribution from orbital moments, even though from
the present measurements we cannot determine its origin,

The moments determined for the 15K data (Figure~\ref{fig1}) and given
in Table~\ref{tab1} show that no significant difference was observed
between the 15K and 40K profiles. Hence, we observe no evidence of the
re-ordering of the Sm moment between these temperatures proposed by
Sampathkumaran {\it et al.} \cite{sampa:96}.

\begin{figure} 
\epsfxsize=220pt
\epsffile{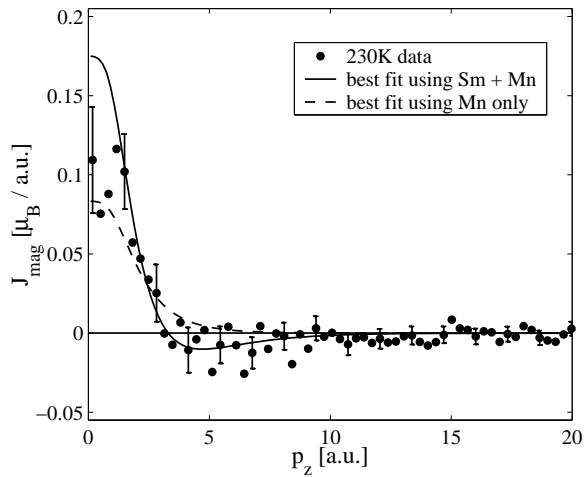} 
\hspace{1cm}
\caption{The experimental magnetic Compton profile of $\rm
SmMn_{2}Ge_{2}$ at $T = 230$K. The lines represent the fitted RHF
profiles obtained using the least squares method for Mn 3$d$ electrons
only and both Sm 4$f$ and Mn 3$d$ electrons. These fits clearly
indicate the presence of a Sm 4$f$ contribution.
\label{fig2}
}
\end{figure}

In Figure~\ref{fig2} we present preliminary MCP data measured at
T=230K. Despite the poorer statistical accuracy it is clear that the
data are still negative for momenta above 3a.u., in fact the area in
the region $4 < p_{z} < 10$a.u. is $-0.04 \pm 0.01 \mu_{B}$. The Sm
4$f$ spin moment is reduced compared with the low temperature
data. The lines shown in Figure~\ref{fig2} represent fits to the data
points first assuming there is no Sm spin moment present (dashed line)
and then assuming contributions from both Sm 4$f$ and Mn 3$d$
electrons (solid line). The least squares fit was made in the momentum
range $1.5 < p_z < 15$ a.u., i.e. beyond the low momentum region where
solid state effects reduce the Mn moment. The curve obtained using Mn
only clearly does not represent a good fit to the data, whereas the
data points show a normal distribution about the fit including both Sm
and Mn contributions, which yields a Sm spin moment of $0.7 \pm 0.1
\mu_{B}$, again aligned antiparallel to the total and Mn 3$d$
magnetizations. As far as the authors are aware, this is the first
time that a Sm spin moment has been observed conclusively in the high
temperature phase in this material. A vanishing Sm moment at high
temperature could not be excluded in the neutron data of Tomka {\it et
al.} \cite{tomka:98} since the refinement procedure was consistent
with moment values ranging from zero to $0.6\mu_{B}$.  It should also
be noted that since the {\it total} moment, measured by neutrons, is
much smaller than the spin and orbital contributions, the magnetic
Compton scattering experiment is more sensitive to the existence of
any ferromagnetic ordering. Furthermore, the MCP result does not rely
on complex data analysis; it can be deduced directly from the raw
data.  The negative tail of the MCPs shown in Figures~\ref{fig1} and
\ref{fig2} cannot be interpreted without a negative 4$f$ contribution
to the moment. 

The bulk magnetization at 230K and 0.96T was measured to be
2.1$\mu_B$. Together with the spin component measured by MCP,
0.7$\mu_B$, this means that there has to be an orbital magnetic moment
of 1.4$\mu_B$. Where this unexpectedly large orbital moment originates
is subject of ongoing research.

In conclusion, our results show that in the low temperature
ferromagnetic phase there is a large spin moment of $3.4 \pm 0.1
\mu_{B}$, negatively polarized with respect to the Mn spin moment and
total magnetization. This, together with the magnetization data, means
that there is also a large orbital moment of similar size. Our data do
not support the existence of a magnetic phase transition near
T=30K. At high temperature, a Sm 4$f$ moment definitely exists, albeit
reduced in size from the low temperature value.

We would like to thank the ESRF for provision of beam time, and the
EPSRC (UK) for financial support. We also thank Clemens
Ritter for critical reading of the manuscript and helpful discussions,
and Martin Lees for performing magnetization measurements at Warwick.
The Ames Laboratory is operated for the US DOE under contract number
W-7405-ENG-82.  This work was supported by the Directorate for Energy
Research, Office of Basic Energy Sciences.


\begin{table}[h]
\caption{\label{tab1} The magnetic moments associated with Sm and Mn
at 15K and 40K.}

\begin{tabular}{lr@{}lr@{}l}
Moments $[\mu_B /\mbox{formula unit}]$& 
\multicolumn{2}{c}{15K} & 
\multicolumn{2}{c}{40K} \\
\hline 
Total spin            & $-$0&.01(3) & $+$0&.02(3) \\ 
Sm spin               & $-$3&.5(1)  & $-$3&.4(1)  \\
Mn + delocalised spin & $+$3&.5(1)  & $+$3&.4(1)  \\ 
\hline
Total magnetisation   & $+$4&.1     & $+$4&.1     \\
\hline
Total orbital         & $+$4&.1     & $+$4&.1     \\ 
\end{tabular}

\end{table}


\begin{thebibliography}{9} 

\bibitem[*]{JAD} Present Address: H.H. Wills Physics Laboratory,
University of Bristol, Tyndall Avenue, Bristol BS8 1TL, UK

\bibitem{szytula:94} A. Szytula and J. Leciejewicz, {\em Handbook of
Crystal Structures and Magnetic Properties of Rare Earth
Intermetallics} (CRC Press, Boca Raton, 1994), pp 114-192.

\bibitem{szytula:89} A. Szytula and J. Leciejewicz, in {\em Handbook
on the Physics and Chemistry of Rare Earths}, edited by
K. A. Gschneidner, Jr. and L. Eyring (Elsevier Science, Amsterdam,
1989), Vol. 12, pp. 133-211.

\bibitem{sampa:96}  E.V. Sampathkumaran, P.L. Paulose and R. Mallik,
\prb {\bf 54}, R3710 (1996).

\bibitem{fujii:85}  H. Fujii, T. Okamoto, T. Shigeoka and N. Iwata,
Solid State Commun. {\bf 53} 715 (1985).

\bibitem{gyorgy:87}  E.M. Gyorgy, B. Batlogg, J.P. Remeika, R.B. van
Dover, R.M. Fleming, H.E. Bair, G.P. Espinosa, A.S. Cooper and R.G.
Maines, J. Appl. Phys {\bf 61}, 4237 (1987).

\bibitem{dover:93}  R.B. van Dover, E.M. Gyorgy, R.J. Cava, J.J.
Krajewski, R.J. Felder and W.F. Peck, \prb {\bf 47}, 6134
(1993).

\bibitem{lord:96}  J.S. Lord, P.C. Riedi, G.J. Tomka, Cz. Kapusta and
K.H.J. Buschow, \prb {\bf 53}, 283 (1996).

\bibitem{tomka:97}  G.J. Tomka, Cz. Kapusta, C. Ritter, P.C. Riedi, R.
Cywinski and K.H.J. Buschow, Physica B {\bf 230-232}, 727 (1997).

\bibitem{tomka:98}  G.J. Tomka, C. Ritter, P.C. Riedi, Cz. Kapusta and
W.  Kocemba, \prb {\bf 58}, 6330 (1998).

\bibitem{platzman:70}  P.M. Platzman and N. Tzoar, \prb {\bf
2}, 3556 (1970).

\bibitem{holm:88}  P. Holm, \pra {\bf 37}, 3706 (1988).

\bibitem{bell:96}  F. Bell, J. Felsteiner and L.P. Pitaevskii,
\pra {\bf 53}, R1213 (1996).

\bibitem{sakai:96}  N. Sakai, J. Appl. Cryst. {\bf 29}, 81 (1996).

\bibitem{cooper:97}  M.J. Cooper, J. Rad. Phys. Chem. {\bf 50}, 63
(1997).

\bibitem{cooper:92}  M.J. Cooper, E. Zukowski, S.P. Collins,
D.N. Timms, F. Itoh and H. Sakurai, J.Phys.:Condens. Matt. {\bf 4},
L399 (1992).

\bibitem{lovesey:96}  S.W. Lovesey, J.Phys.: Condens. Matter {\bf 8},
L353 (1996).

\bibitem{carra:96}  P. Carra, M. Fabrizio, G. Santoro and B.T. Thole,
\prb {\bf 53}, R5994 (1996).

\bibitem{duffy:98}  J.A. Duffy, J.E. McCarthy, S.B. Dugdale, V.
Honkim\"aki, M.J. Cooper, M.A. Alam, T. Jarlborg and S.B. Palmer, J.
Phys.: Condens. Matter {\bf 10}, 10391 (1998).

\bibitem{canfield:92} P.C. Canfield and Z. Fisk,  Phil. Mag. B {\bf
65}, 3151 (1992)

\bibitem{zuko:93} E. Zukowski, S. P. Collins, M.J. Cooper, D.N. Timms,
F. Itoh, H. Sakurai, H. Kawata, Y. Tanaka and A. Maliowski, J. Phys.:
Condens. Matter {\bf 5}, 4077 (1993)

\bibitem{biggs:75} F. Biggs, L.B. Mendelsohn and J.B. Mann, At. Data
Nucl. Data Tables {\bf 16}, 201 (1975).

\bibitem{cooper:96} M.J. Cooper, P.K. Lawson, M.A.G. Dixon,
E. Zukowski, D.N. Timms, F. Itoh, H. Sakurai, H. Kawata, Y. Tanaka and
M. Ito, \prb {\bf 54}, 4068 (1996)

\bibitem{lawson:97} P.K. Lawson, M.J. Cooper, M.A.G. Dixon,
D.N. Timms, E. Zukowski, F. Itoh and H. Sakurai, \prb {\bf 56},
3239 (1997)

\bibitem{dixon:98} M.A.G. Dixon, J.A. Duffy, S. Gardelis,
J.E. McCarthy, M.J. Cooper, S.B. Dugdale, T. Jarlborg and D.N. Timms,
J. Phys.: Condens. Matter {\bf 10}, 2759 (1998).

\bibitem{Fernandez} V. Fernandez, C. Vettier, F. de Bergevin,
C. Giles, and W. Neubeck, \prb {\bf 57}, 7870 (1998); W. Neubeck,
C. Vettier, V. Fernandez, F. de Bergevin, and C. Giles,
J. Appl. Phys. {\bf 85}, 4847 (1999).

\end{thebibliography}
\end{document}